\documentclass[%
twocolumn,
 superscriptaddress,
 amsmath,amssymb,
 aps,
 prx,
]{revtex4-1}

\usepackage{graphicx}
\usepackage{dcolumn}
\usepackage{bm}
\usepackage{isotope}
\usepackage{color}
\usepackage[normalem]{ulem}

\renewcommand{\sout}{\bgroup \color{red} \ULdepth=-.5ex \ULset}

\begin{document}

\preprint{APS/123-QED}

\title{Nuclear liquid-gas phase transition with machine learning}

\author{Rui Wang}
\email{wangrui@sinap.ac.cn}
\affiliation{Key Laboratory of Nuclear Physics and Ion-beam Application~(MOE), Institute of Modern Physics, Fudan University, Shanghai $200433$, China}
\affiliation{Shanghai Institute of Applied Physics, Chinese Academy of Sciences, Shanghai $201800$, China}
\author{Yu-Gang Ma}%
\email{mayugang@fudan.edu.cn}
\affiliation{Key Laboratory of Nuclear Physics and Ion-beam Application~(MOE), Institute of Modern Physics, Fudan University, Shanghai $200433$, China}
\affiliation{Shanghai Institute of Applied Physics, Chinese Academy of Sciences, Shanghai $201800$, China}


\author{R. Wada}
\affiliation{Cyclotron Institute, Texas A$\&$M University, College Station, Texas $77843$, USA}

\author{Lie-Wen Chen}
\affiliation{School of Physics and Astronomy and Shanghai Key Laboratory for Particle Physics and Cosmology, Shanghai Jiao Tong University, Shanghai $200240$, China}

\author{Wan-Bing He}%
\affiliation{Key Laboratory of Nuclear Physics and Ion-beam Application~(MOE), Institute of Modern Physics, Fudan University, Shanghai $200433$, China}

\author{Huan-Ling Liu}
\affiliation{Shanghai Institute of Applied Physics, Chinese Academy of Sciences, Shanghai $201800$, China}

\author{Kai-Jia Sun}
\affiliation{Cyclotron Institute, Texas A$\&$M University, College Station, Texas $77843$, USA}
\affiliation{Department of Physics and Astronomy, Texas A$\&$M University, College Station, Texas $77843$, USA}


\date{\today}

\begin{abstract}
The machine-learning techniques have shown their capability for studying phase transitions in condensed matter physics.
Here, we employ the machine-learning techniques to study the nuclear liquid-gas phase transition.
We adopt an unsupervised learning and classify the liquid and gas phases of nuclei directly from the final state raw experimental data of heavy-ion reactions.
Based on a confusion scheme which combines the supervised and unsupervised learning, we obtain the limiting temperature of the nuclear liquid-gas phase transition.
Its value $9.24\pm0.04~\rm MeV$ is consistent with that obtained by the traditional caloric curve method.
Our study explores the paradigm of combining the machine-learning techniques with heavy-ion experimental data, and it is also instructive for studying the phase transition of other uncontrollable systems, like QCD matter.


\end{abstract}

\maketitle


\section{Introduction}
The nuclear liquid-gas phase transition is an old and long-last topic~\cite{FinPRL49,SieNt305,PanPRL52,RicPR350,MCWPPNP99,BorPPNP105}.
Since the interaction between nucleons exhibit Van der Waals features similar with that between molecules, i.e., a short-distance repulsive core and a long-distance attractive tail, the nuclei, considered as self-bound Fermi liquid, can experience liquid-gas phase transition as well.
Over the past several decades, the nuclear liquid-gas phase transition has been studied based on the heavy-ion collisions at intermediate and relativistic energies and hadron–nucleus collisions at relativistic energies. 
The information of the reaction products are obtained with powerful multidetectors allowing the detection of a large amount of the fragments and light particles produced in the reaction. 
A lot of probes by analyzing sophisticatedly the information of the reaction products have been proposed to recognize the liquid-gas phase transition of nuclei~\cite{PocPRL75,MYGPLB390,MYGPRL83,BotPRL86,EllPRL88,NatPRL89,NatPRC65,LopPRL95,MYGPRC71,BonPRL101,XJPLB727,DXGPRC94,MalPRC95,LHLPRC99}.

The ability of machine-learning techniques~\cite{LeCNt521,JorSc349} of recognizing and characterizing complex sets of data stimulates their applications on physics, and brings new possibilities to the study of the nuclear liquid-gas phase transition.
Besides the common uses like particle identification and tagging in experiments~\cite{BalNtC5,OliJHEP2016,KasJHEP2017}, machine-learning techniques have various novel applications.
Several examples are solving quantum many-body problem~\cite{CarSc355}, analyzing strong gravitational lenses~\cite{HezNt548}, exploring phase properties of quark matter~\cite{PLGNtC9,SteJHEP2019,DYLEPJC80}, constraining and studying field theories~\cite{BrePRL121,ZKPRD100}, and quantum state tomography~\cite{TorNtP14}.
Most notably, in condensed matter physics, machine-learning techniques have been used to classify phases of matter, and identify topological order and phase transitions~\cite{CarNtP13,NieNtP13,chnPRX7,RodNtP15}.

The resemblance between condensed matter physics and nuclear physics is remarkable.
They both involve large amount of degrees of freedom, and share the same theoretical tools like Hartree-Fock theory and finite-temperature field theory.
Experimentally however, unlike the condensed matter physics, the nucleus is an uncontrollable system, and its thermal properties can only be accessed through nuclear reactions.
Thus to treat the nuclear liquid-gas phase transition in terms of equilibrium thermodynamics is not practicable.
The nuclear liquid-gas phase transition is realized through tracing the effect of the spinodal instability, which is intimately related to first-order phase transition, on the reaction dynamics, e.g., by measuring the properties of the intermediate mass fragments~(with charge number larger than $3$).

\begin{figure}[!htb]
\centering
\includegraphics[width=6.5cm]{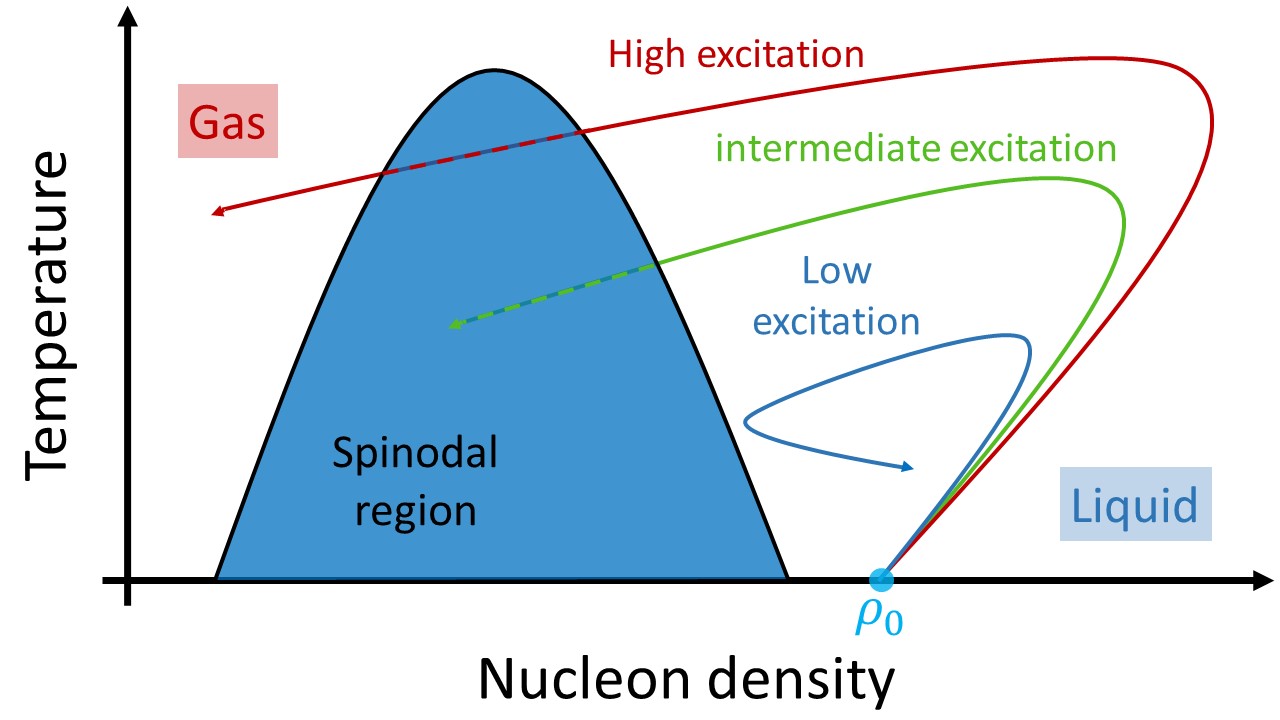}
\caption{
\small The sketch of the phase diagram of nuclear matter and typical phase trajectories of the projectile nucleus in the heavy-ion reactions with different excitation energies.
}
\label{F:SR}
\end{figure}

Heavy-ion reaction experiments may bring the excited nuclei into the spinodal region of the phase diagram in which the spinodal instability may develop exponentially and lead to the break-up of nuclei.
This is commonly referred to as the nuclear multifragmentation.
In Fig.~\ref{F:SR}, we draw schematically the phase diagram and typical phase trajectories of an excited nucleus in heavy-ion reactions.
At the beginning of the reaction, the vast majority of a ground state nucleus is around the saturation density $\rho_0$, which is approximately $0.16$ nucleon $\rm{fm}^{-3}$.
After hitting the target nucleus, the projectile nucleus is excited and compressed, and is regarded as heated liquid~(same for the target nucleus but we will focus on the projectile, which is easier to measure for the detectors).
For low excitation energy, the compressed projectile nucleus will expand and then exhibit a damped monopole oscillation accompanied by the emissions of a few light particles, and remains as liquid.
As the excitation energy increases, the expansion of the nucleus becomes severer and drives the nucleus into the spinodal region.
In the spinodal region, due to the attractive part in the nucleon-nucleon interaction, the high density region will attract nucleons from low density region.
This causes the formation of intermediate mass fragments.
The spinodal decomposition process thus drives the system towards thermodynamic liquid-gas phase coexistence.
If the excitation energy continues to increase, the excited nucleus will expand quickly enough to pass through the spinodal region.
The formation of the intermediate mass fragments becomes less important, since dynamically it takes time for the fragments to format.
In this case, the entire projectile nucleus ends up with dominant light fragments, which corresponds to a gas phase.
Based on the above discussion, the observation of the intermediate mass fragments or nuclear multifragmentation is a sign of the nuclear liquid-gas phase transition.

In the present article, we want to demonstrate that the machine learning techniques are also capable of dealing with the phase transition in realistic nuclear system, other than theoretical models in condensed matter physics, though the way of studying the phase transition of the two systems exhibits essential differences.
To that end, we train the neural network instead of, e.g., by the spin configurations from Monte Carlo simulations, but by the final state information of heavy-ion reaction experiment.
We then use the trained networks to classify the liquid and gas phases of nuclei, and determine the limiting temperature of the nuclear liquid-gas phase transition.

\section{Experimental data}
The reactions of \isotope[40]{Ar} on \isotope[27]{Al} and \isotope[48]{Ti} at $47~\rm MeV/nucleon$ were performed using beams from the TAMU K$500$ super-conducting cyclotron.
The charge and momentum of charged fragments are probed with the $4\pi$ detector, NIMROD~\cite{NIMROD}~(Neutron Ion Multidetector for Reaction Oriented Dynamics).
Generally speaking, phase transitions of small system~(about $10$ constituents) are still well defined, distinguishable~\cite{LabPRL65}, and even detectable~\cite{BluNt334}.
The spectator matter in the nuclear reaction is argued to be ideally suited to investigate the nuclear liquid-gas phase transition since it can largely avoid the effect of dynamical evolution.
Because of the essential binary nature of such reactions, we applied a reconstruction method for the quasi-projectile~(QP) source developed by Ma {\it et al.}~\cite{MYGPRC71}, which perform a three-source (i.e., a QP source, an intermediate velocity source and a quasi-target source) fit for light particles with $Z$ $\leqslant$ $3$, and then use the probability of QP particles to identify the QP light fragments in an event-by-event basis.
For heavier fragments, they can be assigned to the QP source directly through a rapidity cut.
The QP fragments are supposed to come from the excited projectile nucleus.
By the above QP-labeled light and heavier fragments, the mass and charge numbers of QP source could be reconstructed in each event and its excitation energy and other bulk properties can be obtained.
Details of the Ma's QP reconstruction method can be found in Ref.~\cite{MYGPRC71}.
Due to the existence of an intermediate velocity source, the total charge of QP fragments $Z^{\rm QP}$ is generally less than the charge number of the projectile nucleus.
We choose events with $Z^{\rm QP}$ $\geqslant$ $12$ as good QP events, and focus on those $Z^{\rm QP}$ $=$ $12$ events since they have the largest statistics~(using events with other $Z^{\rm QP}$ does not change our conclusion qualitatively).
The universality of the QP fragment distributions~\cite{MYGPRC71} indicates the memory of the entrance channel dynamics is lost prior to the decay of the excited spectators.
Thus the final state information of the QP fragments of the above two reactions is combined to form a single event-by-event data set.
This leads to $40081$ valid QP events with $Z^{\rm QP}$ $=$ $12$.

\begin{figure}[!htb]
\centering
\includegraphics[width=4.5cm]{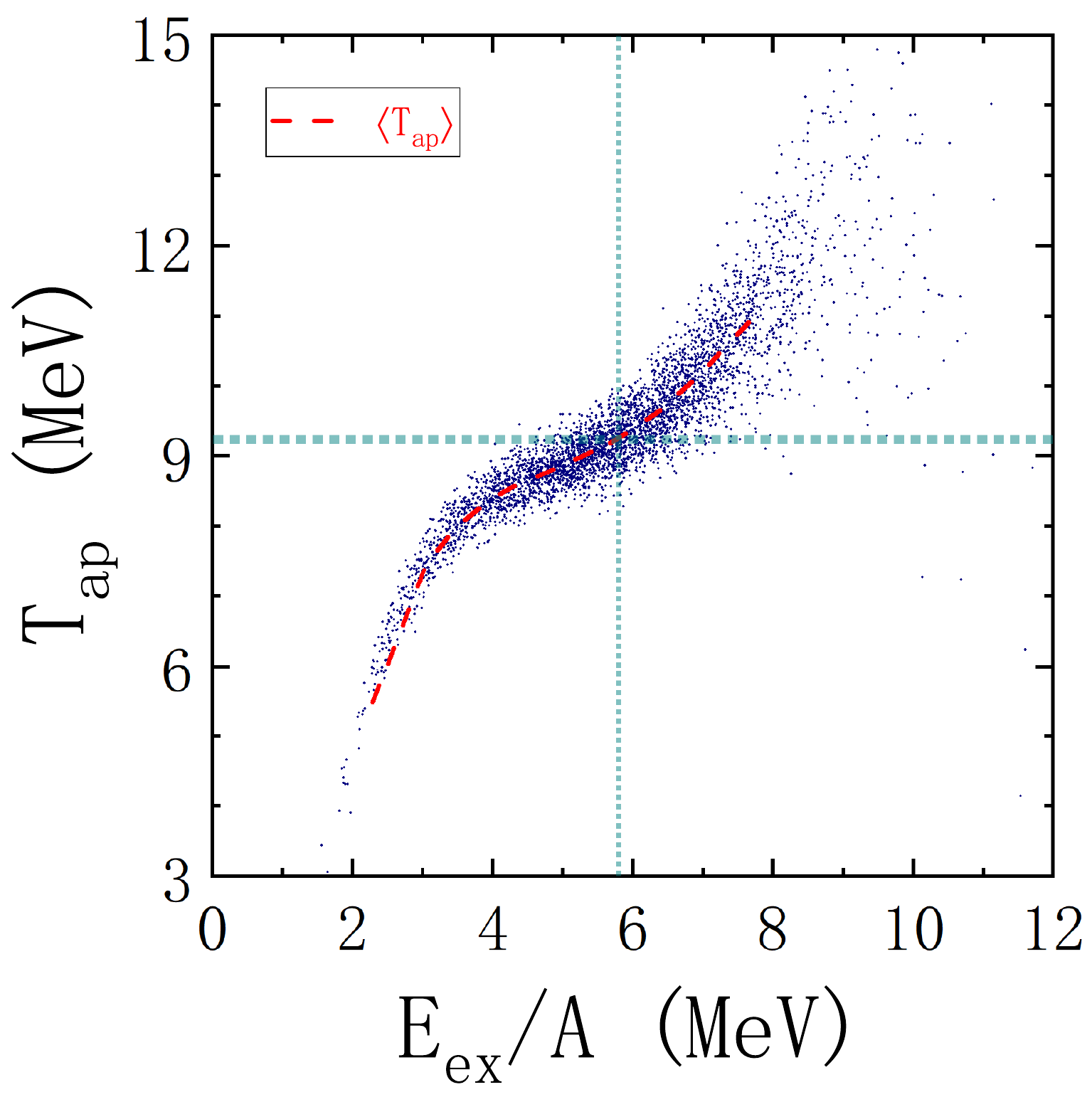}
\caption{\small Scatter plot of the apparent temperature verse the excitation energy per nucleon, only $10\%$ of the total QP events with $Z^{\rm QP}$ $=$ $12$ are shown.
The red dashed line represents $\langle T_{\rm ap}\rangle$ as a function of $E_{\rm ex}/A$.
The horizontal and vertical cyan dotted lines represent the limiting temperature and an analogical characteristic value of $E_{\rm ex}/A$ obtained through confusion scheme, respectively (see below).}
\label{F:CC}
\end{figure}

\begin{figure*}[htb]
\centering
\includegraphics[width=16cm]{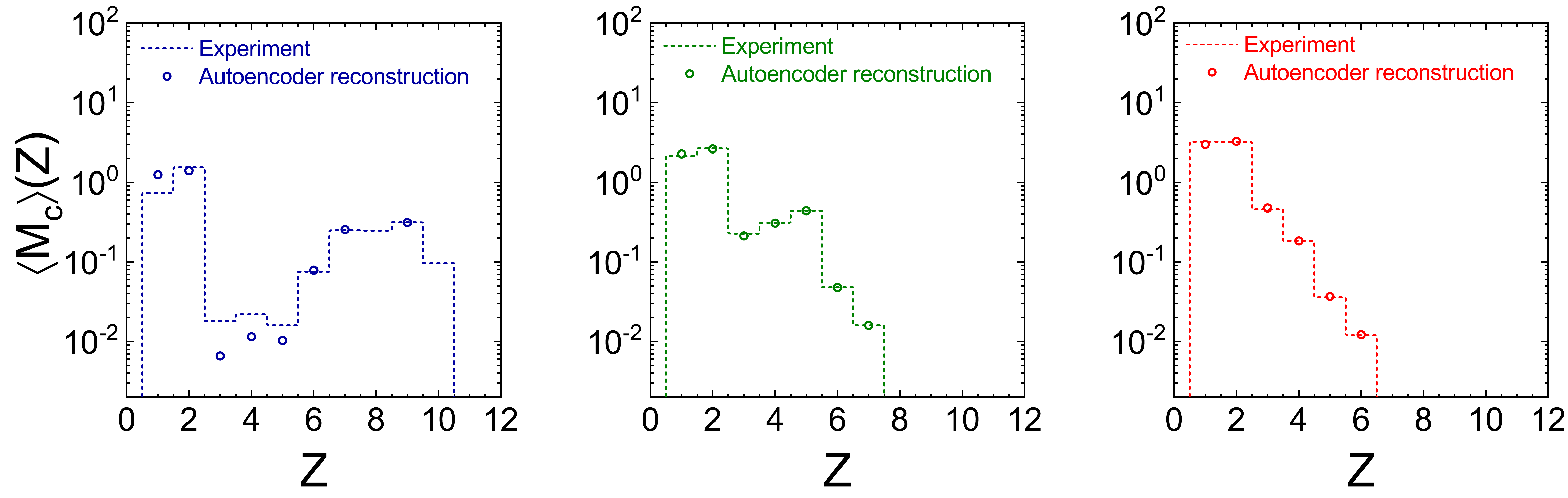}
\put(-380,83){\includegraphics[width=1.5cm]{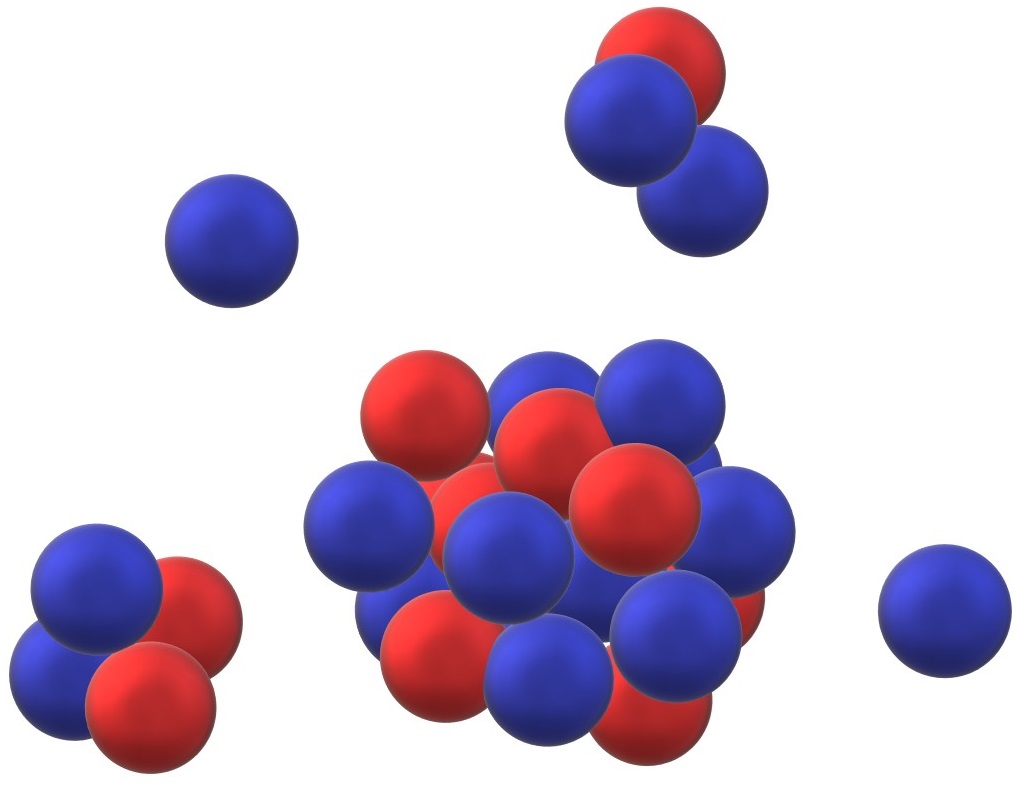}}
\put(-210,70){\includegraphics[width=1.5cm]{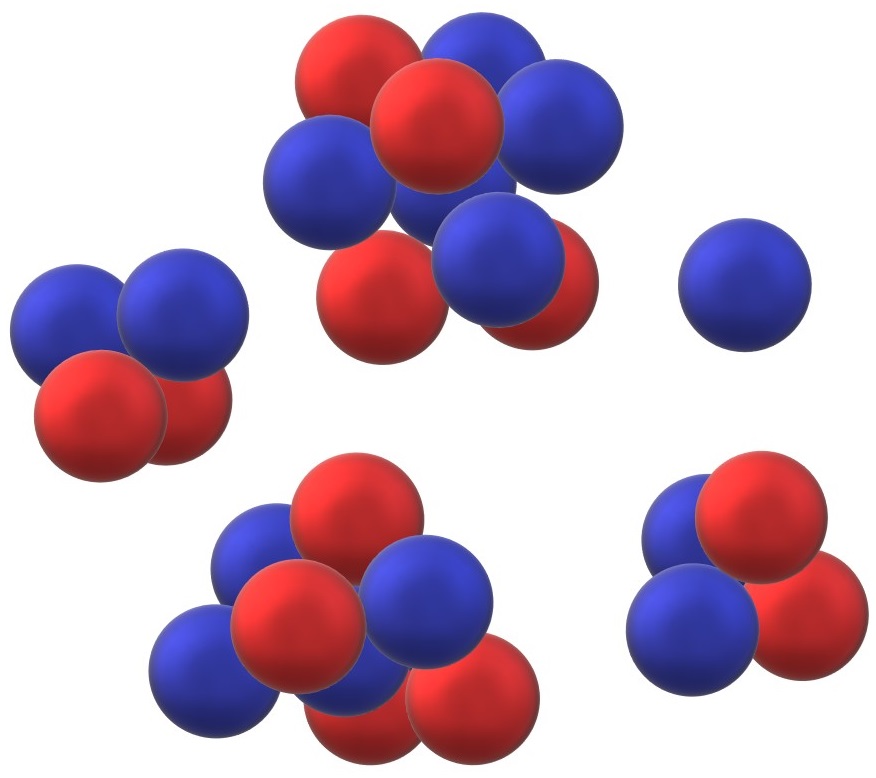}}
\put(-60,65){\includegraphics[width=1.7cm]{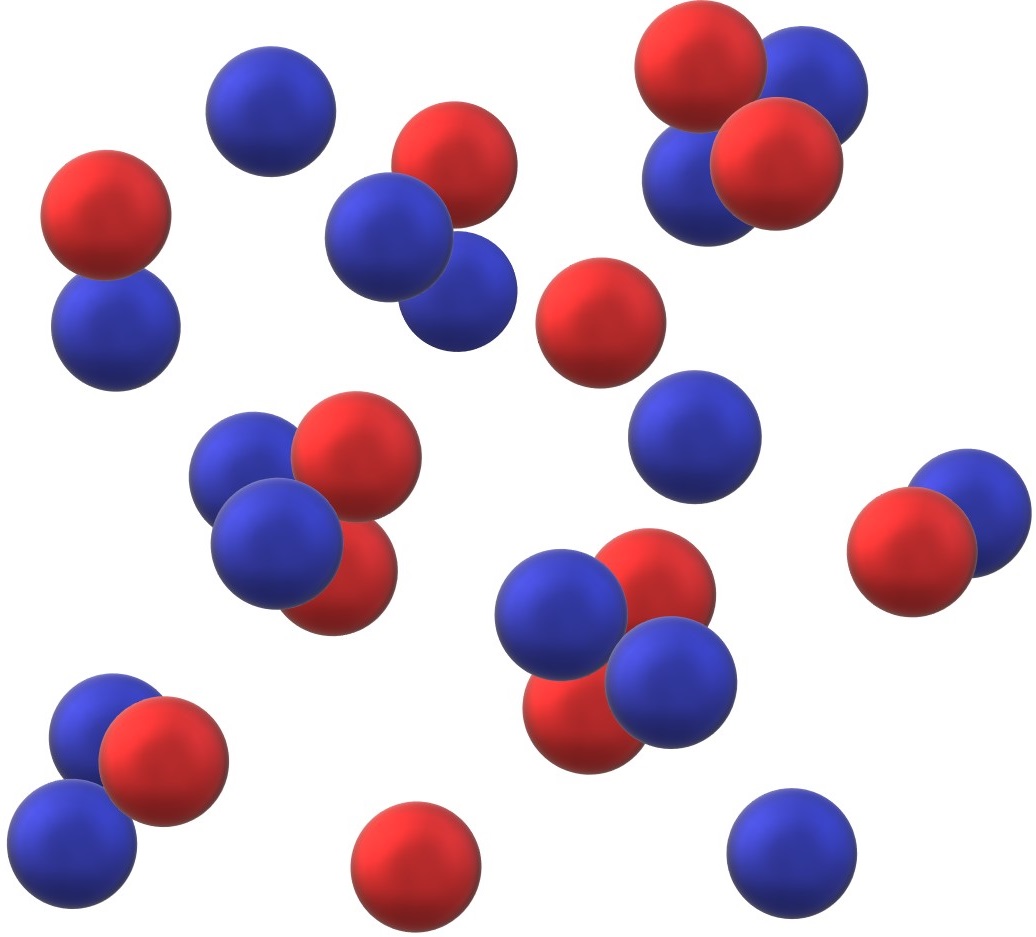}}
\put(-452,136){\bfseries (a)}
\put(-299,136){\bfseries (b)}
\put(-146,136){\bfseries (c)}
\caption{\small The averaged charge multiplicity distribution $\langle M_{\rm c} \rangle(Z)$ of the QP fragments.
The average is taken for different $E_{\rm ex}/A$ bins, left panel for low excitation~($0.9~\rm MeV$ - $2.8~\rm MeV$), middle panel for intermediate excitation~($5.3~\rm MeV$ - $5.4~\rm MeV$), and right panel for high excitation~($8.1~\rm MeV$ - $13.0~\rm MeV$).
The dashed curves represent $\langle M_{\rm c}\rangle(Z)$ from the NIMROD experiment, while the circles from the autoencoder network reconstruction $\langle M_{\rm c}'\rangle(Z)$.
Each $E_{\rm ex}/A$ bin contains $500$ testing events.}
\label{F:CM}
\end{figure*}

Physically, the excitation energy per nucleon $E_{\rm ex}/A$ and apparent temperature $T_{\rm ap}$ of the QP nucleus are used to characterize each QP event.
$E_{\rm ex}$ is deduced event-by-event through \cite{MYGPLB390,MYGPRC71}
\begin{equation}\label{E:E}
    E_{\rm ex} = \sum_{i=1}^{M_{\rm QP}}E_i^{\rm kin} + \frac{3}{2}M_nT - Q.
\end{equation}
The three terms represent the kinetic energy of the charged QP fragments and neutrons, and the mass excess, respectively.
The apparent temperature of the QP nucleus can be obtained by measuring the light particles evaporated from its surface.
This can be achieved via different thermometers, e.g., particle kinetic energy or isotope yield ratios~\cite{TsaPRL78}.
In the present work, the quadruple momentum fluctuation~\cite{WueNPA843} is used as the thermometer.
The quadruple momentum is defined as $Q_{xy} = p_x^2 - p_y^2$, where $p_x$ and $p_y$ are the transverse components of the emitted particle momentum in lab frame.
Since the apparent temperature is derived from the fluctuation of $Q_{xy}$, determining the reaction plane is not necessary.
When the momenta distribute in a Maxwellian form, the average temperature of the events in a given $E_{\rm ex}/A$ bin is related to the variance of $Q_{xy}$, i.e.,

\begin{equation}\label{E:T}
    \langle T_{\rm ap}\rangle = \sqrt{\frac{\langle Q_{xy}^2\rangle - \langle Q_{xy}\rangle^2}{4m^2}},
\end{equation}
where $m$ represents the mass of the probe particle~(deuteron in the present work).
The event-by-event $T_{\rm ap}$ is obtained through a Monte Carlo method based on the standard deviation of $\langle T_{\rm ap}\rangle$~(details can be found in the appendix).
We draw the scatter plot of $T_{\rm ap}$ verse $E_{\rm ex}/A$, or the caloric curve, of the events with $Z^{\rm QP}$ $=$ $12$ in Fig.~\ref{F:CC}.
The red dashed curve in the figure represents the $\langle T_{\rm ap}\rangle$ as a function of $E_{\rm ex}/A$.

\section{Results}
The power of the machine-learning techniques lies in their ability to classify the phases of matter prior to the knowledge of the characteristic quantities, namely, $E_{\rm ex}/A$ and $T_{\rm ap}$.
In that sense, the event-by-event charge-weighted charge multiplicity distribution of QP fragments $ZM_{\rm c}(Z)$ from experiment is used as the input to train the neural network~(the charge weighting is for normalization).
Among the $40081$ valid QP events with $Z^{\rm QP}$ $=$ $12$, $2/3$ are used for training and others for testing.
To obtain an intuitive impression of $M_{\rm c}(Z)$, we average for testing events their $M_{\rm c}(Z)$ in three typical $E_{\rm ex}/A$ bins, namely, low excitation~($0.9$ - $2.8~\rm MeV$), intermediate excitation~($5.3$ - $5.4~\rm MeV$) and high excitation~($8.1$ - $13.0~\rm MeV$), and show $\langle M_{\rm c}\rangle(Z)$ in Fig.~\ref{F:CM} with dashed lines~(each bin contains $500$ events).
The patterns of $\langle M_{\rm c}\rangle(Z)$ reflect the mechanism of the nuclear liquid-gas phase transition discussed above, i.e., a large fragment accompanied by several small fragments at low excitation energy, intermediate mass fragments show up as the excitation energy increases, and the projectile nucleus breaks into small fragments entirely if the excitation energy is high enough.

\subsection{Classifying the liquid and gas phases by the autoencoder method}
\label{S:AE}

\begin{figure}[htb]
\centering
\includegraphics[width=7.0cm]{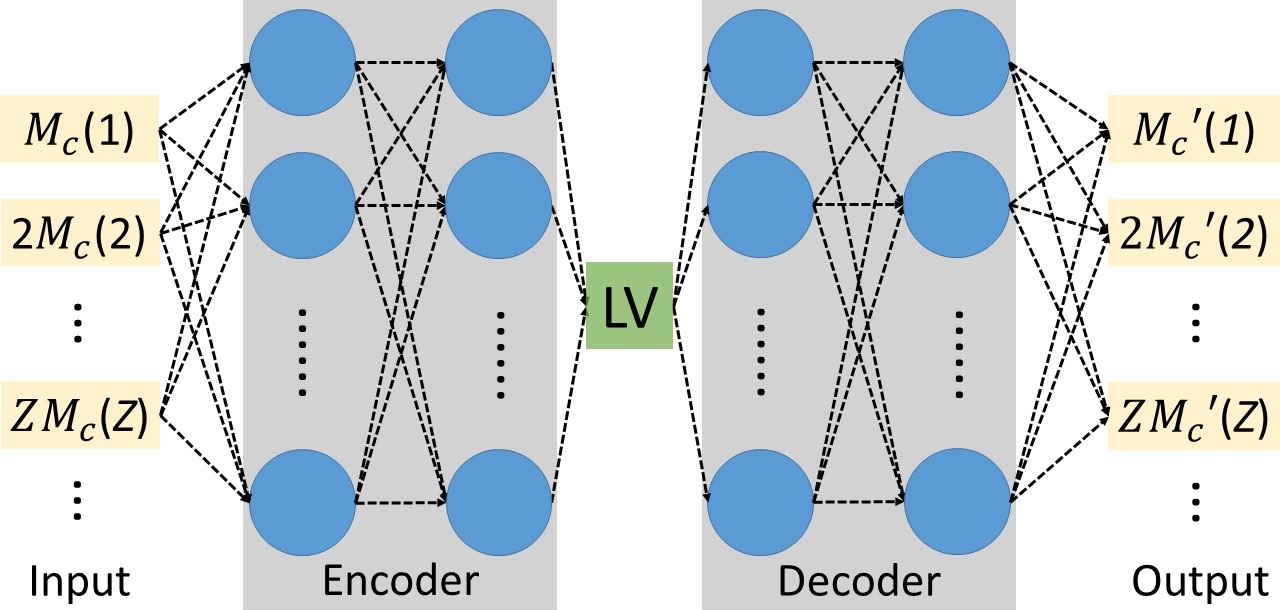}
\caption{\small The construction of the autoencoder network.
We use two layers in the encoder part and decoder part respectively.
The information of the input $ZM_{\rm c}(Z)$ is encoded into the latent variable~(LV) through training the network to best restore the encoded information.
Details of the network is shown in appendix.}
\label{F:AE}
\end{figure}

We first adopt an unsupervised learning, the autoencoder method~\cite{BouBC59}, to study the nuclear liquid-gas phase transition.
We show the construction of the autoencoder network used in the present work in Fig.~\ref{F:AE}.
The neural network consists of two main parts, the encoder part encodes the inputted event-by-event $ZM_{\rm c}(Z)$ to a \emph{latent variable}~(or \emph{code}~\cite{AE-wiki}), and the decoder part decodes the latent variable to $ZM'_{\rm c}(Z)$, and tries to restore the original $ZM_{\rm c}(Z)$.
The neural network is trained to best restore the encoded information, which means the network is trained to minimize the difference between $ZM_{\rm c}(Z)$ and $ZM'_{\rm c}(Z)$.
There are two layers in the encoder and decoder parts respectively, and all layer are fully connected, more details can be found in the appendix.

\begin{figure}[htb]
\includegraphics[width=8.5cm]{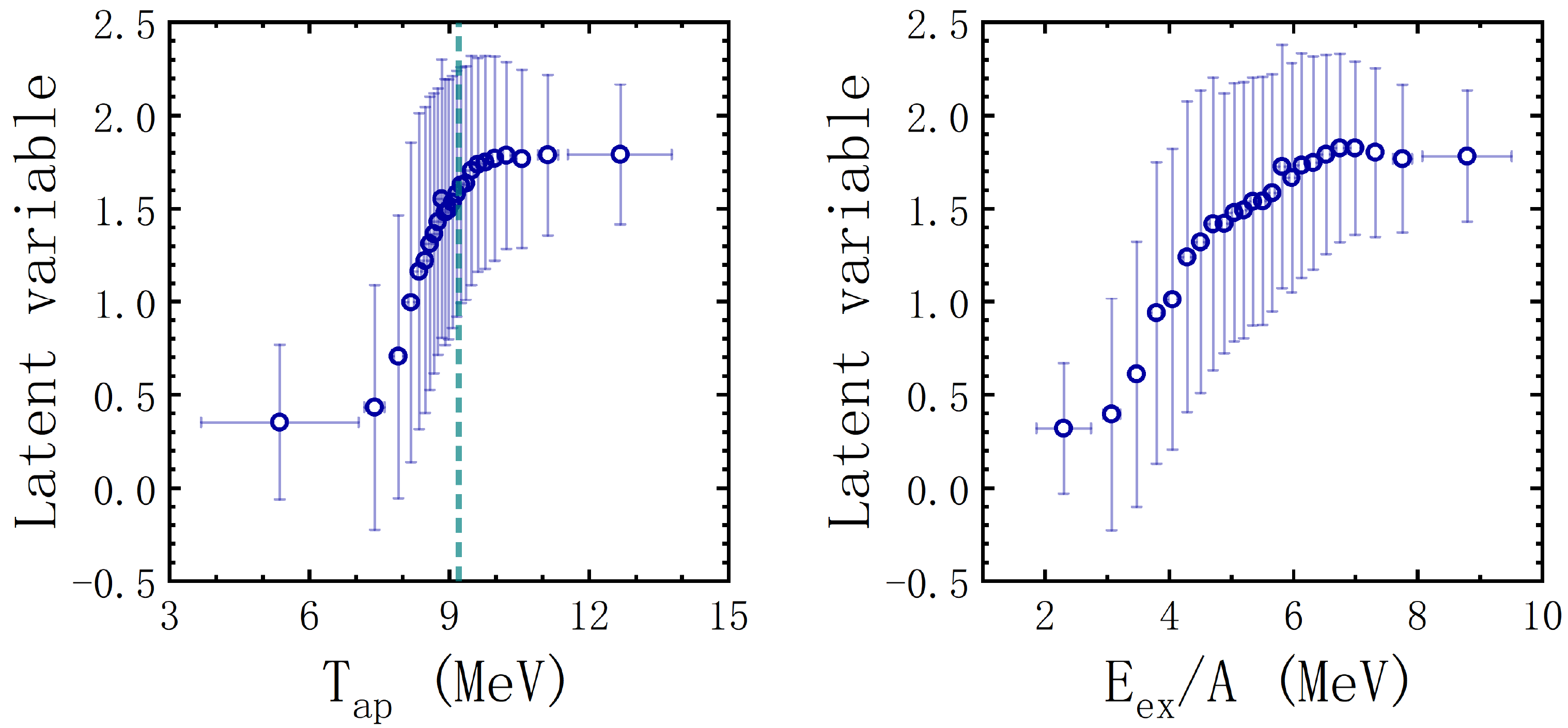}
\put(-147,29){\bfseries (a)}
\put(-22,29){\bfseries (b)}
\caption{\small The mean and standard deviation of the latent variable in different $T_{\rm ap}$ and $E_{\rm ex}/A$ bins, respectively.
Each $T_{\rm ap}$ or $E_{\rm ex}/A$ bin consists of $500$ testing events.
The horizontal errors represent the standard deviation of $T_{\rm ap}$ or $E_{\rm ex}/A$ in that bin.
The cyan vertical dashed line in the left panel represents the limiting temperature obtained through confusion scheme~(see below).}
\label{F:LV}
\end{figure}

For the testing QP events, the reconstructed $M_{\rm c}'(Z)$ are averaged and compared with the original $M_{\rm c}(Z)$ in Fig.~\ref{F:CM}.
We also show the mean and standard deviation of the event-by-event reconstruction loss in the appendix.
We notice from Fig.~\ref{F:CM} that the autoencoder network succeeds in capturing essential information of the inputted event-by-event $ZM_{\rm c}(Z)$.
Once we finish training the autoencoder network, through the charge multiplicity distribution, each QP event is mapped to a number~(latent variable).
We plot in Fig.~\ref{F:LV} the latent variable as a function of $T_{\rm ap}$ and $E_{\rm ex}/A$, with each data point averaged over $500$ testing events.
The vertical error bars in the figure represent the standard deviations of the latent variable of these events, while the horizontal error bars the standard deviations of characteristic parameter ($T_{\rm ap}$ or $E_{\rm ex}/A$).
Although there are large errors due to the event-by-event fluctuations of the experimental charge multiplicity distribution, the averaged latent variable as a function of $T_{\rm ap}$ or $E_{\rm ex}/A$ exhibits a sigmoid pattern, which indicates the trained autoencoder network treats the low and high temperature~(or low and high excitation energy) regions differently.
Considering that the autoencoder network is trained prior to any knowledge of the characteristic quantities, i.e., $E_{\rm ex}/A$ and $T_{\rm ap}$, the autoencoder network is capable of classifying different phases of nuclei directly from the final state information of the heavy-ion experiment.
The area in the midst of the two phases represents those liquid-gas coexistence events that enter the spinodal region and affected by the spinodal instability.
It is interesting for further studies to find out how the latent variable is related to physical quantities.

\subsection{Limiting temperature from confusion scheme}
Traditionally, the nuclear liquid-gas phase transition can be recognized from the relation between $E_{\rm ex}/A$ and the apparent temperature $T_{\rm ap}$, namely, the caloric curve~\cite{PocPRL75}.
As the excitation energy increases, more energy is transferred to the internal energy of the liquid phase projectile nucleus, and the apparent temperature increases.
Dramatic change happens if the excited projectile nucleus goes into the spinodal region.
Part of the excitation energy is consumed to form the fragments, in other words, transfer to latent heat.
As a consequence, the increase of the apparent temperature slows down significantly, and even a plateau in the caloric curve is observed~\cite{PocPRL75,NatPRC65}.
The specific heat capacity of the collision process $\tilde{c}$ is defined to describe quantitatively the effect of the spinodal instability, i.e., 
\begin{equation}
    \tilde{c} \equiv \frac{d(E_{\rm ex}/A)}{dT_{\rm ap}},
\end{equation}
which can be obtained from the caloric curve shown in Fig.~\ref{F:CC}.
Note its difference with $c_{\rm P}$ and $c_{\rm V}$ since the external conditions on pressure and volume is not practicable due to the complex nature of such finite, self-bound object like nucleus.
The specific heat capacity $\tilde{c}$ will reach a peak when the spinodal instability affects the excited projectile nucleus the most severely.
The corresponding apparent temperature at the maximum is called limiting temperature, which can be used to deduce the critical temperature of infinite nuclear matter~\cite{NatPRL89}.
Although the limiting temperature is different from the first-order critical temperature of isobaric process, the non-monotonic structure of $\tilde{c}(T_{\rm ap})$ indicates the existence of the spinodal region in nuclear matter phase diagram, which is a convincing evidence of the existence of nuclear liquid-gas~(first-order) phase transition.

The above picture of the nuclear liquid-gas phase transition is consistent with the feature of the latent variable shown in Fig.~\ref{F:LV}, and we can further obtain the limiting temperature by a confusion scheme~\cite{NieNtP13}.
In the confusion scheme, the neural network is trained with data that are deliberately labelled incorrectly according to a proposed critical point, and the phase transition properties can be deduced from the performance curve, i.e., the total testing accuracy as a function of the proposed critical point, of the neural network~\cite{NieNtP13}.
In the original confusion scheme~\cite{NieNtP13}, a W-shape performance curve of the Ising model is observed, and the proposed critical point corresponding to the local maximum in the middle of the curve is recognized as the realistic critical point.
As mentioned above, the nucleus is an uncontrollable system, thus the events used to train the neural network can not come from separate phases like those in the Ising model.
The change from liquid to gas phase is gradual through a liquid-gas phase coexistence.
Therefore as we will demonstrate below, a W-shape performance curve in the case of the Ising model, is replaced by a V-shape, when adopting the confusion scheme in the nuclear liquid-gas phase transition.

\begin{figure}[htb]
\centering
\includegraphics[width=5.0cm]{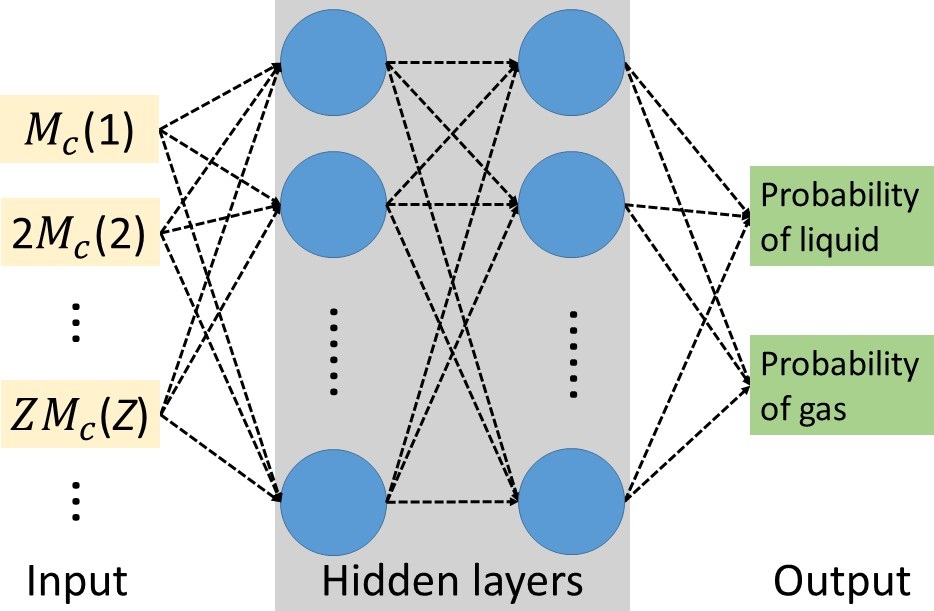}
\caption{\small The construction of the Bayesian neural network for supervised learning.
The network consists of two hidden layers, details are shown in the appendix.}
\label{F:CS}
\end{figure}

\begin{figure}[htb]
\includegraphics[width=8.5cm]{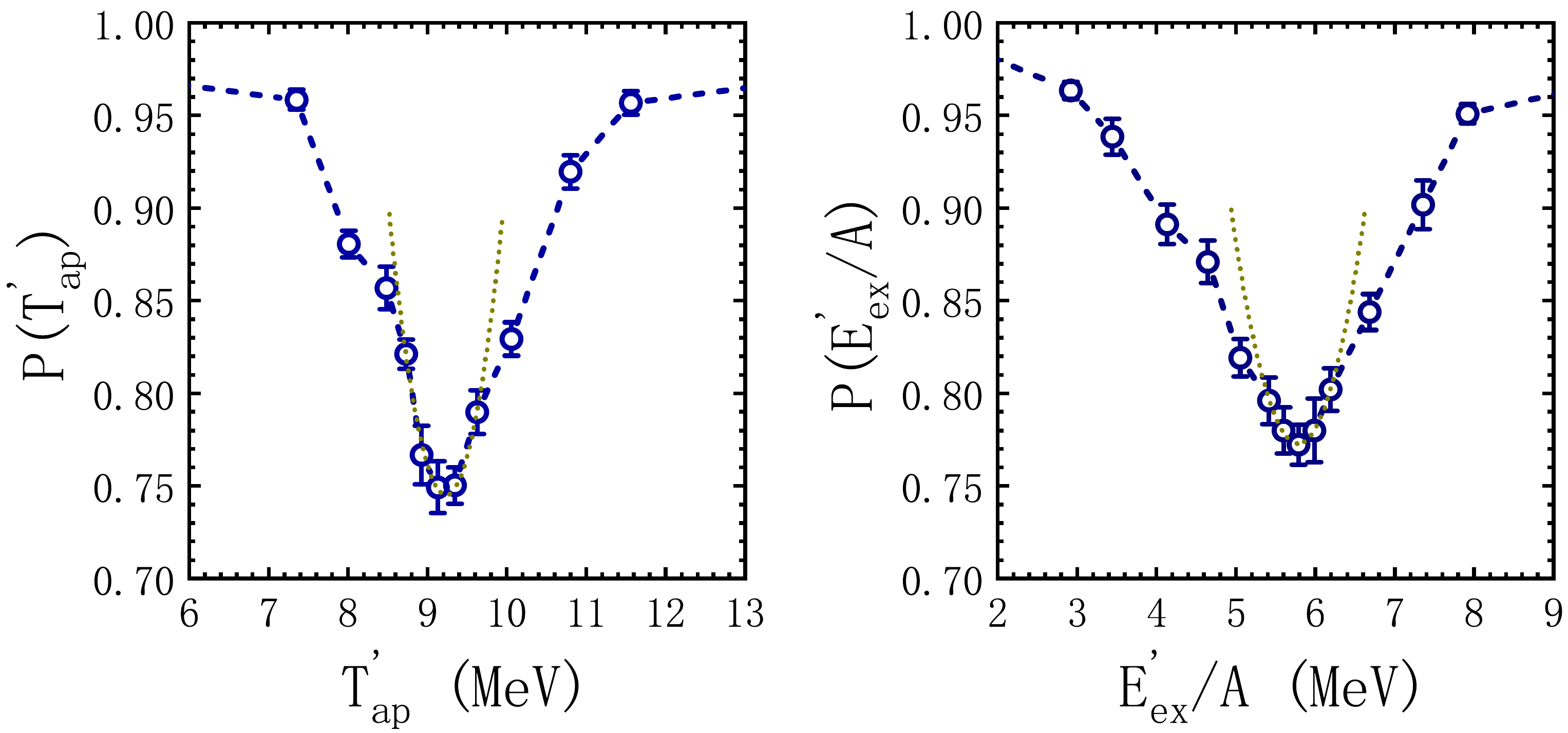}
\put(-143,29){\bfseries (a)}
\put(-20,29){\bfseries (b)}
\caption{\small The performance curve $P(T_{\rm ap}')$ and $P(E_{\rm ex}'/A)$, i.e., the testing accuracy as a function of the proposed temperature $T_{\rm ap}'$ and transition excitation energy $E_{\rm ex}'$, respectively.
The yellow dotted lines represent a parabolic fit of the lowest five data points with errors.}
\label{F:PC}
\end{figure}

Taking the example of $T_{\rm ap}$, the picture is as follows.
In order to properly include the uncertainty of the obtained limiting temperature, we construct a Bayesian neural network~(BNN) which contains two fully connected layers, as shown in Fig.~\ref{F:CS}, to perform this supervised classification.
The QP events are divided into two categories and labelled as liquid-like or gas-like according to a proposed transition temperature $T_{\rm ap}'$.
When $T_{\rm ap}'$ increases from low temperature~(vice versa when $T_{\rm ap}'$ decreases from high temperature), the feature of the liquid phase~(a large fragment accompanied by several small fragments) emerges on both the liquid-like and gas-like category.
As a consequence, the testing accuracy of low temperature region below $T_{\rm ap}'$ is relatively low, and the total performance of the network $P(T_{\rm ap}')$ starts to decrease, as shown in the left panel of Fig.~\ref{F:PC}.
When $T_{\rm ap}'$ continues to increase, the features of the intermediate mass fragments begin to show up in the liquid-like category and the total performance continues to decrease.
As $T_{\rm ap}'$ approaches the realistic limiting temperature, the features of the intermediate mass fragments become evident in both the low and high temperature categories, thus significantly reduces the total efficiency of the neural network.
The total testing accuracy then reaches its minimum at $T_{\rm ap}'$ $\approx$ $T_{\rm lim}$.
The error bars in the figure are the standard deviation of the testing accuracy, and are obtained based on the trained BNN by performing $10$ test runs, with each consists $1000$ random selected testing events.
The limiting temperature is obtained by a parabolic fit of the lowest five data points with errors.
The limiting temperature through the confusion scheme is $9.24\pm0.04~\rm MeV$, which is consistent with the $9.0\pm0.4~\rm MeV$ obtained from the traditional analysis of caloric curve~\cite{WadPRC99}.
Besides that, it also locates in the intermediate temperature region in the left panel of Fig.~\ref{F:LV}~(cyan vertical dashed line), which indicates both the autoencoder network and confusion scheme learn the basic feature of liquid and gas phase from the raw event-by-event charge multiplicity distribution.

In the right panel of Fig.~\ref{F:PC}, similar analysis is carried out on $E_{\rm ex}/A$, and we obtain an analogical characteristic value of $E_{\rm ex}/A$ $=$ $5.79\pm0.02~\rm MeV$.
Since the network classifies each event purely based on $ZM_{\rm c}(Z)$, the liquid-like events and gas-like events divided by the characteristic value of $T_{\rm ap}$~(horizontal dotted cyan line in Fig.~\ref{F:CC}) correspond to almost same events that divided by the characteristic value of $E_{\rm ex}/A$~(vertical dotted cyan line in Fig.~\ref{F:CC}).
Considering that the latent variable obtained in Sec.~\ref{S:AE} is correlated with $T_{\rm ap}$ or $E_{\rm ex}/A$, performing the above analysis with the latent variable will leads to similar result.

\section{summary and outlook}
We have shown that the machine-learning techniques can be employed to a traditional nuclear physics topic, the nuclear liquid-gas phase transition.
Based on the experiment event-by-event charge multiplicity distribution, the neural networks are capable of classifying the liquid and gas phases, and determining the limiting temperature of the nuclear liquid-gas phase transition.
The hidden parameters identified by machine learning may acquire physical significance.
The latent variable can be related to the order parameters for certain systems~\cite{HWPRE95,WetPRE96}.
A new field called 'softness', which characterizes the local structure, is used to study the correlations between structure and dynamics in glassy liquids~\cite{SchNtP12}.
To relate the latent variable in Fig.~\ref{F:LV} to certain physical quantities might be meaningful for future studies.
The analysis performed here can also be applied to study the first-order phase transition of the QCD matter by choosing proper final state observables, since the picture of the first-order phase transition of the QCD matter is quite similar with that of the nuclear matter~\cite{SKJPLB781}.
We anticipate more sophisticated observables like the kinetic energy spectra of the final state particles, along with more advanced machine-learning techniques will provide us new features in the nuclear liquid-gas phase transition, even in other fields of nuclear physics, that beyond the present knowledge.

\begin{acknowledgments}
We thank Meisen Gao, Jie Pu and Ying Zhou for the maintenance of the GPU severs, Xiaopeng Zhang for technical support, and Feng Li for useful discussion.
This work is partially supported by the National Natural Science Foundation of China under Contracts No. $11890714$, No. $11421505$ and No. $11625521$, the Key Research Program of Frontier Sciences of the CAS under Grant No. QYZDJ-SSW-SLH$002$, the Strategic Priority Research Program of the CAS under Grants No. XDPB$09$ and No. XDB$16$, the US Department of Energy under Grant No. DE-FG$02$-$93$ER$40773$ and No. de-sc$0015266$, and the Robert A. Welch Foundation under Grant A$330$ and No. A-$1358$.
\end{acknowledgments}

\appendix

\section{Event-by-event excitation energy}
The event-by-event excitation energy is obtained through Eq.~(\ref{E:E}).
In Eq.~(\ref{E:E}), $E_i^{\rm kin}$ represents the kinetic energy of the $i$-th charged QP fragment, which is obtained through the measured data directly.
The contribution of the neutron kinetic energy is taken as $3/2M_nT$.
Since the NIMROD does not provide the momentum of neutrons, the associated QP neutrons can not be separated through the three source fit.
Therefore the neutron multiplicity $M_n$ is approximated as the difference between the assumed total QP mass and the sum of the detected masses of the QP fragments, i.e., $M_n$ $=$ $A^{\rm QP}$ $-$ $\sum A_i^{\rm QP}$.
The total mass of the QP spectator $A^{\rm QP}$ is determined from $Z^{\rm QP}$~(the sum of $Z_i^{\rm QP}$, with $Z_i^{\rm QP}$ being the charge of the $i$-th measured fragments), by assuming the QP spectator has the same neutron-proton ratio as the initial projectile.
The temperature $T$ is assumed to equal to the temperature of QP protons which is taken from the three source fit parameters.
The non-observed light charged particles from the QP spectator are calculated from the extracted three source fit parameters and added to the QP in mass and energy.

\section{Event-by-event apparent temperature}
The apparent temperature of the QP nucleus is obtained via the quadrupole momentum fluctuation.
For the QP events in a given excitation energy bin, the experimental deuteron quadrupole momentum fluctuation temperature $\langle T_{\rm ap}\rangle$ is obtained through Eq.~(\ref{E:T}), and its standard deviation values $\Delta\langle T_{\rm ap}\rangle$ are evaluated by the TProfile class of ROOT in the CERN data analysis library~\cite{Root}.
Both $\langle T_{\rm ap}\rangle$ and $\Delta\langle T_{\rm ap}\rangle$ are fitted by polynomial functions to obtain their $E_{\rm ex}/A$ dependence, i.e., $T_{\rm pol}(E_{\rm ex}/A)$ and $\Delta T_{\rm pol}(E_{\rm ex}/A)$.
Since the standard deviation $\Delta\langle T_{\rm ap}\rangle$ is small, the event-by-event apparent temperature of the reconstructed QP nucleus is evaluated through
\begin{equation}
    T_{\rm ap} = T_{\rm pol}(E_{\rm ex}/A) + \Delta T_{\rm pol}(E_{\rm ex}/A){\cal N}(0,1),
\end{equation}
where ${\cal N}(0,1)$ is a zero-mean Gaussian random number with unit variance.

\section{Details of the neural network}
The neural network contains successively one input layer, several hidden layers, and one output layer.
Each layer provides its output $\textbf{z}$ through a matrix multiplication of its input $\textbf{x}$, i.e., $\textbf{z}$ = $\textbf{W}\cdot\textbf{x}$ $+$ $\textbf{b}$.
The elements in the matrix $\textbf{W}$ are known as weights and in the vector $\textbf{b}$ as biases.
In a normal full-connected neural network these parameters are single values, while in a Bayesian neural network~(BNN) they become distributions, usually assumed to be Gaussian form.
The layer is then followed by an activation function, $f(\textbf{z})$, which turns a linear transform to a non-linear one.
Commonly used activation functions are \emph{sigmoid}, \emph{tanh}, and \emph{ReLU}~(rectified linear unit).
$f(\textbf{z})$ is then used as the input of the next layer.
The neural network can be treated as a function $\tilde{\textbf{y}}$ $=$ $g(\textbf{x};\textbf{W},\textbf{b})$, which transfers non-linearly a given input $\textbf{x}$ to an output predictions $\tilde{\textbf{y}}$.
The cost function of the network $C(\tilde{\textbf{y}},\textbf{y})$ is used to measure the difference between the network predictions $\tilde{\textbf{y}}$ and their true values $\textbf{y}$.
The neural network is trained to minimize $C(\tilde{\textbf{y}},\textbf{y})$, by adjusting its parameters $\textbf{W}$ and $\textbf{b}$.

\begin{figure}[htb]
\centering
\includegraphics[width=8.5cm]{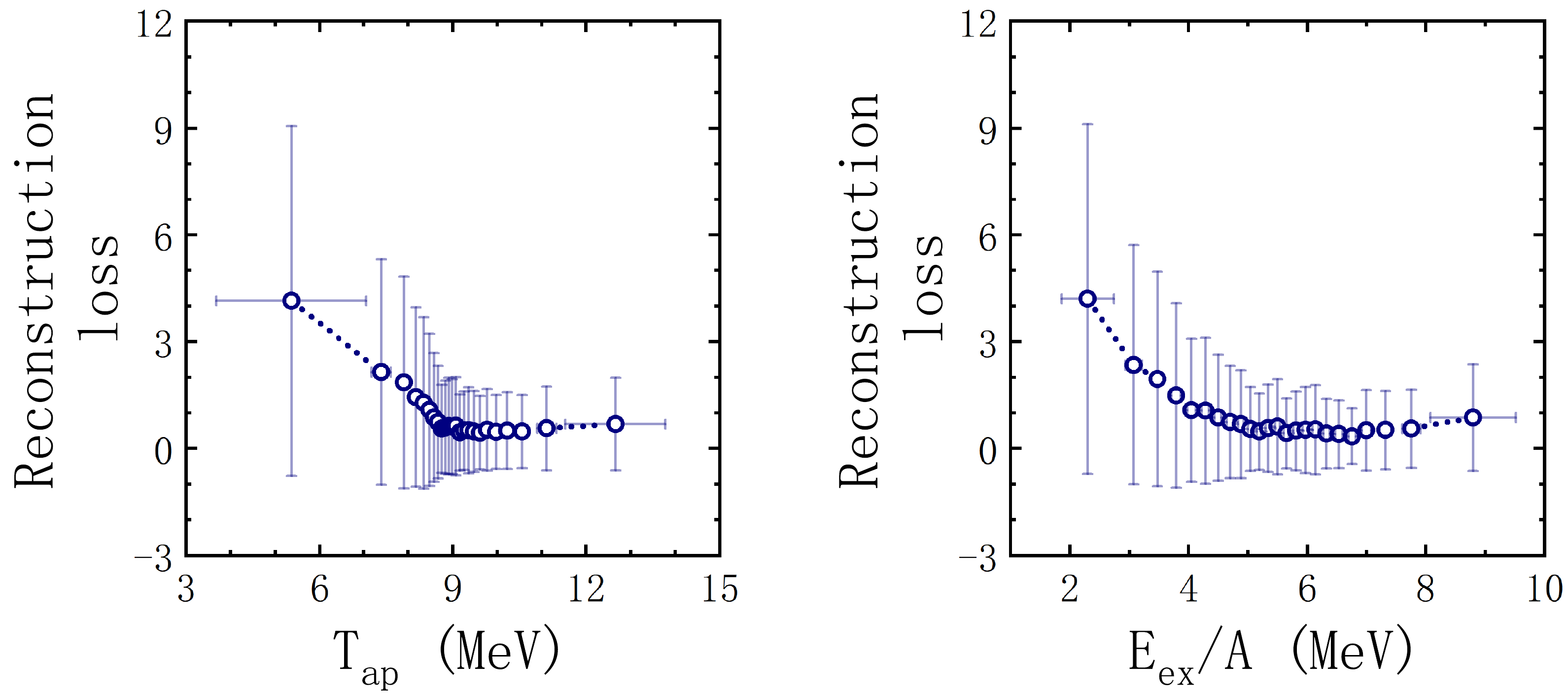}
\put(-148,90){\bfseries (a)}
\put(-20,90){\bfseries (b)}
\caption{\small The mean and standard deviation of the event-by-event reconstruction loss of the auto-encoder network for different a) $T_{\rm ap}$ and b) $E_{\rm ex}/A$ bins, respectively.
Each bin contains $500$ testing events.
The horizontal errors represent the standard deviation of $T_{\rm ap}$ or $E_{\rm ex}/A$ in that bin.}
\label{F:loss}
\end{figure}

For the autoencoder network used in the present work shown in Fig.~\ref{F:AE}, the optimization is fulfilled by the $Adam$~\cite{Adam} package in \emph{Tensorflow} for this full-connected network.
When training the network, we use an exponential decreasing learning rate $\alpha$ $=$ $10^{-3}$ $+$ $(10^{-3} - 10^{-6})\exp(-i/10000)$, with $i$ the training epoch.
The cost function is defined as $C(\tilde{\textbf{y}},\textbf{y})$ $=$ $(\tilde{\textbf{y}}$ $-$ $\textbf{y})^2$.
To prevent the network from over fitting the data, we adopt a standard $l_2$ regularization term in the cost function of this neural network.
The cost function is adjusted to include the norm of the weight $\textbf{W}$ and the bias $\textbf{b}$, i.e., $\tilde{C}(\tilde{\textbf{y}},\textbf{y})$ $=$ $C(\tilde{\textbf{y}},\textbf{y})$ $+$ $l_2(\Vert\textbf{W}\Vert^2/2 + \Vert\textbf{b}\Vert^2/2)$, with $l_2$ a positive number.
The $l_2$ regularization prevents the weights and biases from increasing to arbitrary large values during the optimization.
We list the information of the autoencoder network in Table~\ref{T:N}.
The encoder part and decoder part are mirror symmetric and each consists of two layers. 
Since the output layer represents the positive defined reconstructed charge-weighted charge multiplicity distribution $ZM_{\rm c}'(Z)$, a \emph{ReLU} is used to connect the decoder and the output layer.
Figs.~\ref{F:CM} and \ref{F:LV} in the main text show the result with $8$ and $4$ neurons in the first and second encoder layer, respectively, and \emph{ReLU} activation between encoder part and latent variable~(with regularization $l_2$ $=$ $0.1$).
Note that not all the network constructions shown in the second column of Table~\ref{T:N} restore properly the inputted charge multiplicity distributions.
For those who restore properly the charge multiplicity distributions, their feature of the latent variable shown in Fig.~\ref{F:LV} are quite similar.
As a supplement to Fig.~\ref{F:CM}, we show the mean and standard deviation of the event-by-event reconstruction loss in Fig.~\ref{F:loss}.
The event-by-event reconstruction loss is defined as the sum of absolute deviation of the charge multiplicity distribution, $\sum_{Z = 1}^{12}|ZM_{\rm c}(Z) - ZM'_{\rm c}(Z)|$.
We note that the average reconstruction loss for the event with low temperature is larger than that with higher temperature.
Considering ZMc(Z) is normalized to 12 and the event-by-event nature, we think this is still acceptable.

\begin{table}[!htp]
\centering
\caption{\small Details of the autoencoder network used in the present work.}
\begin{tabular}{ccc}
\hline\hline
 Layer & Neuron number & Activation\\
\hline
 Input & $12$ & \emph{tanh}\\
 Encoder $1$st layer & $8-64$ & \emph{tanh}\\
 Encoder $2$nd layer& $4-32$ & \emph{tanh} or \emph{ReLU}\\
 latent variable & $1$ & \emph{tanh}\\
 decoder $1$st layer & $4-32$ & \emph{tanh}\\
 decoder $2$nd layer & $8-64$ & \emph{ReLU}\\
 Output & $12$ & $-$ \\
\hline\hline
\end{tabular}
\label{T:N}
\end{table}

For the supervised learning BNN used in the present work shown in Fig.~\ref{F:CS}, we use two hidden layers, each consists of 100 neurons.
The input layer and the hidden layers are followed by \emph{ReLU}.
Weights and biases of the neurons are represented as Gaussian distributions.
The maximization of the evidence lower bound (ELBO) is employed to solve the Bayes' formula.
Here we do not employ the $l_2$ regularization since the problem of over-fit is not severe in BNN.



%

\end{document}